\documentclass[journal, 10pt]{IEEEtran}
\usepackage[absolute, showboxes]{textpos}
\TPMargin{5pt}
\newcommand{\copyrightstatement}{
\begin{textblock}{0.80}(0.10, 0.01) 
\noindent
\footnotesize
\copyright 2021 IEEE.
Personal use of this material is permitted.
Permission from IEEE must be obtained for all other uses, in any current or future media, including reprinting/republishing this material for advertising or promotional purposes, creating new collective works, for resale or redistribution to servers or lists, or reuse of any copyrighted component of this work in other works.
\end{textblock}
}
\usepackage{amsmath, amssymb, amsfonts, amsthm, mathtools}
\allowdisplaybreaks
\usepackage{url, bm, cite, color}

\usepackage{multirow, multicol}
\usepackage{caption, subcaption}
\usepackage[ruled,vlined]{algorithm2e}
\SetKw{Continue}{continue}
\usepackage{setspace}
\newlength\mylen

\theoremstyle{definition}

\newtheorem{remark}{Remark}

\DeclareMathOperator*{\trace}{tr}

\DeclareMathOperator*{\argmin}{argmin}

\DeclareMathOperator*{\s}{s}
\DeclareMathOperator*{\z}{z}
\def\BibTeX{{\rm B\kern-.05em{\sc i\kern-.025em b}\kern-.08em T\kern-.1667em\lower.7ex\hbox{E}\kern-.125emX}}

\begin{document}
\copyrightstatement
\title{Independent Vector Extraction for Fast \\Joint Blind Source Separation and Dereverberation}

\author{
    Rintaro~Ikeshita, \IEEEmembership{Member, IEEE},
    and Tomohiro~Nakatani, \IEEEmembership{Fellow, IEEE}%
    \thanks{Manuscript received Feb xx, 2021. (Corresponding author: Rintaro Ikeshita)}
}
\maketitle

\begin{abstract}
We address a blind source separation (BSS) problem in a noisy reverberant environment in which the number of microphones $M$ is greater than the number of sources of interest, and the other noise components can be approximated as stationary and Gaussian distributed.
Conventional BSS algorithms for the optimization of a multi-input multi-output convolutional beamformer have suffered from a huge computational cost when $M$ is large.
We here propose a computationally efficient method that integrates a weighted prediction error (WPE) dereverberation method and a fast BSS method called independent vector extraction (IVE), which has been developed for less reverberant environments.
We show that, given the power spectrum for each source, the optimization problem of the new method can be reduced to that of IVE by exploiting the stationary condition, which makes the optimization easy to handle and computationally efficient.
An experiment of speech signal separation shows that, compared to a conventional method that integrates WPE and independent vector analysis, our proposed method achieves much faster convergence while maintaining its separation performance.
\end{abstract}

\IEEEpeerreviewmaketitle

\section{Introduction}
\label{sec:intro}

When multiple speech signals are observed by distant microphones (e.g., in a conference room),
they are contaminated with reverberation and background noise.
The problem of extracting each speech signal and removing the reverberation and background noise from only the observed signal
is called (convolutive) blind source separation or extraction
(BSE)~\cite{pedersen2008convolutive,comon2010handbook,cichocki2002adaptive}.
Here, we consider BSE in the short-term Fourier transform (STFT) domain
under the following two conditions:
\begin{itemize}
\item The reverberation time ($\mathrm{RT}_{60}$) is larger than the frame length of the STFT, and the mixture should be treated as a convolutive mixture in the STFT domain as well.
%
\item The number of microphones $M$ is greater than that of speech signals $K$
and there can be background noise.
\end{itemize}

To cope with reverberation,
one can apply a dereverberation method~\cite{naylor2010speech} such as weighted prediction error (WPE)~\cite{nakatani2010wpe,yoshioka2011wpe-ica,yoshioka2012wpe-mimo}
as preprocessing of BSE for instantaneous mixtures in the STFT domain (called BSE-inst in this paper).
We then apply some BSE-inst method
such as independent vector analysis (IVA)~\cite{kim2007,hiroe2006,li2009joint}
and independent vector extraction (IVE)~\cite{koldovsky2018ive,jansky2020adaptive,scheibler2019overiva,scheibler2020fast,scheibler2020ive,ike2020ive,ike2020overiva}
developed for less reverberant environments, to extract $K$ speech signals.
Such a cascade configuration of WPE and IVA/IVE has a low computational cost,
but the WPE dereverberation filter is estimated without considering the separation attained by IVA/IVE following WPE.

To jointly optimize the WPE dereverberation and separation filters through a unified optimization,
methods that integrate WPE and several BSE-inst methods have been proposed~\cite{yoshioka2011wpe-ica,yoshioka2012wpe-mimo,nakatani2020computationally,ike2019ilrma,kagami2018wpe-ilrma},
and it has been reported that these methods can give higher separation performance than the cascade configuration of WPE and BSE-inst (see, e.g.,~\cite{nakatani2020computationally}).
However, the computational cost of optimizing both WPE and BSE-inst models becomes huge when $M$ is large.

To reduce the computational cost of the conventional joint optimization methods while maintaining their separation performance, 
we propose a new BSE method called \textit{IVE for convolutive mixtures (IVE-conv)}, which integrates WPE and IVE (Section~\ref{sec:model}).
We show that, given source power spectra,
the IVE-conv optimization problem can be reduced to the IVE optimization problem by exploiting the stationary condition,
and this reduction is not computationally intensive (Section~\ref{sec:reduction}).
The IVE optimization problem can be solved fast~\cite{scheibler2019overiva,scheibler2020fast,scheibler2020ive,ike2020ive,ike2020overiva},
and so can the IVE-conv optimization problem (Section~\ref{sec:alg:1}).
We also propose another new algorithm for IVE-conv that alternately optimizes WPE and IVE (Section~\ref{sec:alg:2}).
Similar algorithms have already been developed in~\cite{yoshioka2011wpe-ica,yoshioka2012wpe-mimo}, but our proposed one significantly reduces the computational time complexity of the conventional ones.
In a numerical experiment in which two speech signals are extracted from mixtures,
we show the effectiveness of our new approach.

\section{Blind source extraction problem}
\label{sec:conv-bse-problem}

Let $M$ be the number of microphones.
Suppose that an observed mixture $\bm{x} \coloneqq \{ \bm{x}(f,t) \}_{f,t} \subset \mathbb{C}^M$ in the STFT domain is a convolutive mixture of
$K$ nonstationary source signals and $N_{\z} \coloneqq M - K$ background noise signals:%
\footnote{
The assumption that the dimension of the noise signal is $M - K$ concerns the rigorous development of efficient algorithms
and can be violated to some extent when applied in practice (see numerical experiments in Section~\ref{sec:exp}).
}
\begin{align}
\nonumber
\bm{x}(f,t) = \sum_{\tau = 0}^{N_{\tau}} \left[
    \sum_{i = 1}^{K} \bm{a}_i (f,\tau) s_i(f,t - \tau) + A_{\z}(f,\tau) \bm{z}(f, t - \tau)
\right],
\end{align}
\vspace{-3 mm}
\begin{alignat}{5}
\label{eq:ai}
& \bm{a}_i(f,\tau) &&\in \mathbb{C}^M,
\quad
& s_i(f,t) &\in \mathbb{C}, \quad i \in \{ 1,\ldots,K \},
\\
\label{eq:Az}
& A_{\z}(f,\tau) &&\in \mathbb{C}^{M \times N_{\z}},
\quad
& \bm{z}(f,t) &\in \mathbb{C}^{N_{\z}}.
\end{alignat}
Here, $f = 1,\ldots,F$ and $t = 1,\ldots,T$ denote the frequency bin and time frame indexes, respectively.
Also, $s_i(f,t) \in \mathbb{C}$ and $\bm{z}(f,t) \in \mathbb{C}^{N_{\z}}$ are the signals of the target source $i = 1,\ldots,K$ and the background noises, respectively.
$\{ \bm{a}_i(f,\tau) \}_{\tau = 0}^{N_\tau}$ and $\{ A_{\z}(f,\tau) \}_{\tau = 0}^{N_\tau}$ are the acoustic transfer functions (ATFs) for the corresponding sources,
where $N_\tau + 1$ is the length of the ATFs.
The BSE problem addressed in this paper is defined as the problem of estimating
the sources of interest, i.e.,
$\{ s_i(f,t) \}_{i,f,t}$.
We assume that $K$ is given and the background noises are more stationary than the sources of interest.

\section{Probabilistic model}
\label{sec:model}

We present the proposed IVE-conv model that integrates WPE~\cite{yoshioka2011wpe-ica,yoshioka2012wpe-mimo,nakatani2010wpe} and IVE~\cite{scheibler2019overiva,scheibler2020fast,scheibler2020ive,ike2020overiva,ike2020ive,koldovsky2018ive,jansky2020adaptive}.
Let $\hat{\bm{x}}(f,t) \in \mathbb{C}^{M + L}$ with $L = M(D_2 - D_1 + 1)$ and $0 < D_1 \leq D_2$ be given by
\begin{align}
\nonumber
\hat{\bm{x}}(f,t) = [\, \bm{x}(f,t)^\top, \bm{x}(f,t - D_1)^\top, \ldots, \bm{x}(f,t - D_2)^\top \,]^\top,
\end{align}
where $\empty^\top$ is the transpose of a vector.
Suppose that there exists a convolutional filter
$\hat{W}(f) \in \mathbb{C}^{(M + L) \times M}$ satisfying
\begin{align}
\label{eq:s=Px}
s_i(f,t) &= \hat{\bm{w}}_i(f)^h \hat{\bm{x}}(f,t) \in \mathbb{C}, \quad i \in \{1,\ldots,K\},
\\
\label{eq:z=Px}
\bm{z}(f,t) &= \hat{W}_{\z}(f)^h \hat{\bm{x}}(f,t) \in \mathbb{C}^{N_{\z}},
\\
\hat{W}(f) &= [\hat{\bm{w}}_1(f), \ldots, \hat{\bm{w}}_{K}(f), \hat{W}_{\z}(f)] \in \mathbb{C}^{(M + L) \times M},
\end{align}
where $\empty^h$ denotes the conjugate transpose.
As pointed out in~\cite{boeddeker2020jointly,nakatani2020cbf}, convolutional filter $\hat{W}(f)$ can be decomposed into
the WPE prediction matrix $G(f) \in \mathbb{C}^{L \times M}$ and the ICA separation matrix $W(f) \in \mathbb{C}^{M \times M}$:
\begin{align}
\label{eq:GW}
\hat{W}(f) &=
\begin{bmatrix}
    W(f) \\
    -G(f) W(f)
\end{bmatrix}
=
\begin{bmatrix}
    I_M \\
    -G(f)
\end{bmatrix}
W(f).
\end{align}
Here, $I_d \in \mathbb{C}^{d \times d}$ is the identity matrix.

We also assume that the original source signals are mutually independent and that 
the target source (resp. noise) signals obey time-dependent (resp. time-independent) complex Gaussian distributions
in the same way as in IVE~\cite{ike2020ive,ike2020overiva,scheibler2020fast,scheibler2020ive,scheibler2019overiva,jansky2020adaptive,koldovsky2018ive}:
\begin{align}
\label{eq:si:vec}
\bm{s}_i(t) &\coloneqq [s_i(1,t), \ldots, s_i(F,t)]^\top \in \mathbb{C}^F,
\\
\label{eq:si:gauss}
\bm{s}_i(t) &\sim \mathbb{C}\mathcal{N} \left( \bm{0}_F, v_i(t) I_F \right), \quad v_i(t) \in \mathbb{R}_{> 0},
\\
\label{eq:z-pdf}
\bm{z}(f,t) &\sim \mathbb{C}\mathcal{N} \left( \bm{0}_{N_{\z}}, \Omega(f) \right),
\quad \Omega(f) \in \mathcal{S}_{++}^{N_{\z}},
\\
& \hspace{-10 mm}
\label{eq:iid}
\text{
    $\{ \bm{s}_i(t), \bm{z}(f,t) \}_{i,f,t}$ are mutually independent.
}
\end{align}
Here, $\bm{0}_d \in \mathbb{C}^d$ is the zero vector,
$\mathcal{S}_{++}^d$ denotes the set of all Hermitian positive definite matrices of size $d \times d$,
and $\mathbb{R}_{> 0} = \mathcal{S}_{++}^1$.
Assumption \eqref{eq:z-pdf} that the background noise signal is stationary and Gaussian distributed is essential for developing computationally efficient algorithms.
In Section~\ref{sec:exp}, we will experimentally show that this assumption can be violated to some extent when applied in practice.

The IVE-conv model is defined by \eqref{eq:s=Px}--\eqref{eq:iid}.
The parameters $\hat{W} \coloneqq \{ \hat{W}(f) \}_f$,
$v \coloneqq \{ v_i(t) \}_{i,t}$,
and $\Omega \coloneqq \{ \Omega(f) \}_f$
can be estimated based on maximum likelihood,
which is equivalent to minimizing $\hat{g}(\hat{W},\Omega,v) \coloneqq - \frac{1}{T} \log p(\bm{x})$:
\begin{align}
\nonumber
\hat{g}(\hat{W}, \Omega, v)
&=
\sum_{f = 1}^F
\sum_{i = 1}^{K} 
\Big[
    \hat{\bm{w}}_i(f)^h \hat{R}_i(f) \hat{\bm{w}}_i(f) + \frac{1}{T} \sum_{t = 1}^T \log v_i(t)
\Big]
\\
\nonumber
&
+ \sum_{f = 1}^F \trace \big( \hat{W}_{\z}(f)^h \hat{R}_{\z}(f) \hat{W}_{\z}(f) \Omega(f)^{-1} \big)
\\
\label{eq:obj}
&
- \sum_{f = 1}^F \log \det \big( W(f)^h W(f) \Omega(f)^{-1} \big),
\\
\nonumber
\hat{R}_i(f) &=
\frac{1}{T} \sum_{t = 1}^T
\frac{\hat{\bm{x}}(f,t) \hat{\bm{x}}(f,t)^h }{v_i(t)}, \quad i \in \{1,\ldots,K,\z\},
\end{align}
where we define $v_{\z}(t) = 1$ for all $t = 1,\ldots,T$
(see, e.g.,~\cite{ike2019ilrma} for the derivation of $\hat{g}$).
If $L = 0$ and $\hat{W}(f) = W(f)$, then objective function $\hat{g}$ has the same form as the counterparts of ICA, IVA, and IVE,
which has been discussed extensively in the literature~\cite{pham2001,degerine2006maxdet,yeredor2012SeDJoCo,ono2010auxica,ono2011auxiva,ike2020ive,ike2020overiva,scheibler2019overiva,scheibler2020fast,scheibler2020ive}.
For $L \geq 1$, $K = M$, and $N_{\z} = 0$, the optimization problem has been discussed explicitly in~\cite{ike2019ilrma,nakatani2020computationally}
and implicitly in~\cite{yoshioka2011wpe-ica,yoshioka2012wpe-mimo,kagami2018wpe-ilrma}.
\begin{remark}
\label{remark:WPE-ICA}
The proposed IVE-conv is an integration of WPE and IVE.
If we replace IVE with ICA, IVA, or independent low-rank matrix analysis (ILRMA)~\cite{kitamura2016ilrma},
then the IVE-conv turns out to be the method that integrates
WPE with ICA~\cite{yoshioka2011wpe-ica},
WPE with IVA (IVA-conv)~\cite{nakatani2020computationally},
or WPE with ILRMA~\cite{kagami2018wpe-ilrma,ike2019ilrma}, respectively.
In this sense, the novelty of the IVE-conv model might seem limited.
However, if $M$ gets large, computationally efficient algorithms can be developed only for IVE-conv,
which is our main contribution.%
\footnote{
This letter is based on our work~\cite{ike2020asj} reported in a domestic workshop in which an algorithm similar to but less efficient than Algorithm 1 (proposed in Section~\ref{sec:alg:1}) was first presented.
Recently, as follow-up research of our previous work~\cite{ike2020asj}, a method has been developed~\cite{togami2020over}
that replaces the IVE-conv spectrum model~\eqref{eq:si:vec}--\eqref{eq:si:gauss} with a model using nonnegative matrix factorization (NMF)~\cite{lee1999nmf,fevotte2009,smaragdis2003NMF}.
In contrast, here, we develop a more efficient Algorithm 1 in a rigorous way by providing new insight into the IVE-conv optimization problem in Section~\ref{sec:reduction}.
In addition, Algorithm 2 proposed in Section~\ref{sec:alg:2} is completely new.
}
\end{remark}
%

%
%

\section{Optimization algorithm}
\label{sec:alg}

\begin{figure}[t]
\begin{center}
\includegraphics[width=0.99\linewidth]{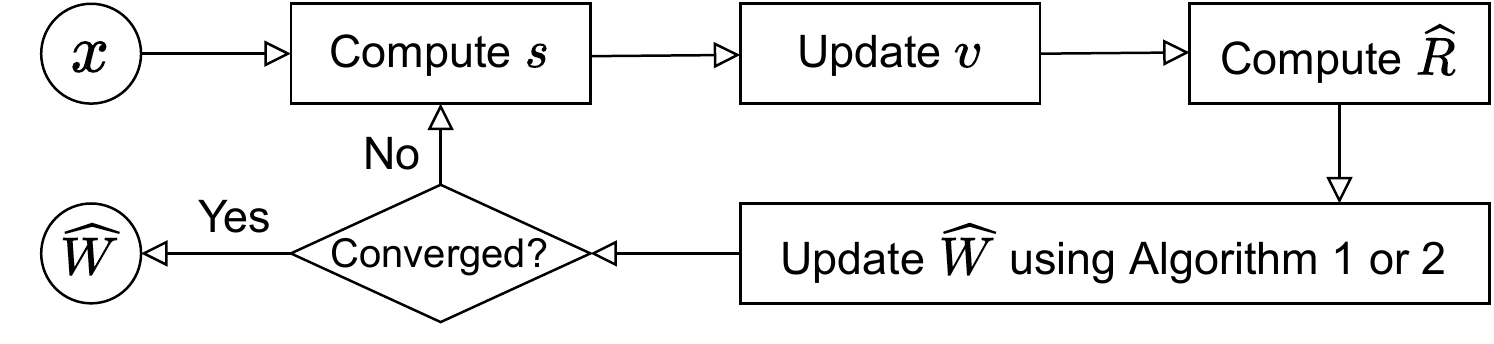}
\end{center}
\vspace{-4 mm}
\caption{
Flowchart of IVE-conv
}
\label{fig:process-flow}
\vspace{-3 mm}
\end{figure}

To obtain a local optimal solution for the minimization problem of \eqref{eq:obj},
two block coordinate descent (BCD~\cite{tseng2001convergence}) algorithms summarized in Table~\ref{table:alg} will be developed.
All the algorithms shown in Table~\ref{table:alg} update $v$ and $(\hat{W}, \Omega)$ alternately.
The flowchart of IVE-conv is shown in Figure~\ref{fig:process-flow}.

When $(\hat{W}, \Omega)$ are kept fixed, $v$ can be optimized as
\begin{align}
v_i(t) = \frac{1}{F} \| \bm{s}_i(t) \|_2^2 = \frac{1}{F} \bm{s}_i(t)^h \bm{s}_i(t).
\end{align}

In what follows, we will develop two BCDs to optimize
$(\hat{W}, \Omega)$ while keeping $v$ fixed.
Because this subproblem can be addressed independently for each frequency bin,
we focus only on optimizing $\hat{W}(f)$ and $\Omega(f)$, and the frequency bin index $f$ is dropped off to ease the notation.
Also, we will denote the submatrices of $\hat{W}$ and $\hat{R}_i$, $i \in \{1,\ldots,K,\z\}$ as
\begin{align}
\label{eq:W}
\hat{W} &=
\begin{bmatrix}
W \\
-G W
\end{bmatrix}
=
\begin{bmatrix}
W \\
\bar{W}
\end{bmatrix} 
=
\left[
\begin{array}{c|c|c|c}
\bm{w}_1 & \cdots & \bm{w}_K & W_{\z}
\\
\bar{\bm{w}}_1 & \cdots & \bar{\bm{w}}_K & \bar{W}_{\z}
\end{array}
\right],
\\
\nonumber
\bm{w}_i &\in \mathbb{C}^{M},
\quad \bar{\bm{w}}_i \in \mathbb{C}^{L},
\quad W_{\z} \in \mathbb{C}^{M \times N_{\z}},
\quad \bar{W}_{\z} \in \mathbb{C}^{L \times N_{\z}},
\\
\nonumber
\hat{R}_i &= \begin{bmatrix}
R_i & \bar{P}_i^h \\
\bar{P}_i & \bar{R}_i
\end{bmatrix} \in \mathcal{S}_{++}^{M + L},
\quad
\bar{P}_i
\in \mathbb{C}^{L \times M},
\quad \bar{R}_i \in \mathcal{S}_{++}^L.
\end{align}

\subsection{Reduction from IVE-conv to IVE when $v$ is kept fixed}
\label{sec:reduction}

Before developing the algorithms, we show that the problem of minimizing $\hat{g}$ with respect to $\hat{W}$ and $\Omega$ (when source power spectra $v = \{v_i(t)\}_{i,t}$ are kept fixed), i.e.,
\begin{align}
\label{problem:maxdet-conv}
(\hat{W}, \Omega) \in \argmin_{(\hat{W},\, \Omega)} \hat{g}(\hat{W}, \Omega, v),
\end{align}
can be reduced to problem \eqref{problem:maxdet} below
that has been addressed in the study of
IVE~\cite{ike2020ive,ike2020overiva,scheibler2019overiva,scheibler2020fast,scheibler2020ive}.

Every optimal $\bar{W}$ (the lower part of $\hat{W}$) in problem~\eqref{problem:maxdet-conv}  satisfies the stationary condition~\cite{nocedal-Jorge2006optimization},
which is computed as
\begin{align}
\nonumber
\frac{\partial \hat{g}}{\partial \bar{\bm{w}}_i^\ast}
&= \bm{0}_L
&~~&\Longleftrightarrow&~~&
\bar{P}_i \bm{w}_i + \bar{R}_i \bar{\bm{w}}_i = \bm{0}_L \in \mathbb{C}^L,
\\
\label{eq:opt:wi-wpe}
&&~~&\Longleftrightarrow&~~&
\bar{\bm{w}}_i = -\bar{R}_i^{-1} \bar{P}_i \bm{w}_i \in \mathbb{C}^L,
\\
\nonumber
\frac{\partial \hat{g}}{\partial \bar{W}_{\z}^\ast}
&= O
&~~&\Longleftrightarrow&~~&
\bar{P}_{\z} W_{\z} + \bar{R}_{\z} \bar{W}_{\z} = O \in \mathbb{C}^{L \times N_{\z}},
\\
\label{eq:opt:Wz-wpe}
&&~~&\Longleftrightarrow&~~&
\bar{W}_{\z} = -\bar{R}_{\z}^{-1} \bar{P}_{\z} W_{\z} \in \mathbb{C}^{L \times N_{\z}},
\end{align}
where $\empty^\ast$ denotes the element-wise conjugate.
Eqs. \eqref{eq:opt:wi-wpe} and \eqref{eq:opt:Wz-wpe} imply that the optimal $\bar{W}$ is a function of $W$
and that the variable $\bar{W}$ can be removed from $\hat{g}$ by substituting \eqref{eq:opt:wi-wpe} and \eqref{eq:opt:Wz-wpe}.
In other words, problem \eqref{problem:maxdet-conv} is equivalent to the following problem through \eqref{eq:opt:wi-wpe} and \eqref{eq:opt:Wz-wpe}: 
\begin{align}
\label{problem:maxdet}
&
(W, \Omega) \in \argmin_{(W,\, \Omega)} g(W, \Omega, v),
\\
\nonumber
g &= 
\sum_{i = 1}^{K} \bm{w}_i^h V_i \bm{w}_i
+ \trace \big( W_{\z}^h V_{\z} W_{\z} \Omega^{-1} \big)
- \log \det \big( W^h W \Omega^{-1} \big),
\\
\label{eq:Vi}
V_i &\coloneqq R_i - \bar{P}_i^h \bar{R}_i^{-1} \bar{P}_i  \in \mathcal{S}_{++}^M, \quad i \in \{ 1,\ldots,K,\z \}.
\end{align}
Since problem \eqref{problem:maxdet} is nothing but the problem
addressed in the study of IVE, we can directly apply efficient algorithms that have been developed for 
IVE~\cite{ike2020ive,ike2020overiva,scheibler2019overiva,scheibler2020fast,scheibler2020ive}.
Our new algorithm developed in Section~\ref{sec:alg:1} is based on this observation.

\begin{table*}[t]
\begin{center}
{\footnotesize
\caption{Optimization process of BCD}
\label{table:alg}
\begin{tabular}{cccll} \hline
& Method & Reference & \multicolumn{1}{c}{Optimization process$\empty^{1)}$} & \multicolumn{1}{c}{Computational time complexity}
\\ \hline
\multirow{2}{*}{Conventional}
& \multirow{2}{*}{IVA-conv$\empty^{2)}$}
& \cite{ike2019ilrma,nakatani2020computationally}
& $v \rightarrow \hat{\bm{w}}_1 \rightarrow \cdots \rightarrow \hat{\bm{w}}_K \rightarrow \hat{\bm{w}}_{K + 1} \rightarrow \cdots \rightarrow \hat{\bm{w}}_M$
& $\mathrm{O} (M L^2 FT + M L^3 F )$
\\
&& \cite{yoshioka2011wpe-ica,yoshioka2012wpe-mimo,kagami2018wpe-ilrma,nakatani2020computationally}
& $v \rightarrow G \rightarrow \bm{w}_1 \rightarrow \cdots \rightarrow \bm{w}_K \rightarrow \bm{w}_{K + 1} \rightarrow \cdots \rightarrow \bm{w}_M$
& $\mathrm{O}( M L^2 FT + M^3 L^3 F )$ in~\cite{nakatani2020computationally}
\\
\hline
\multirow{2}{*}{Proposed}
& \multirow{2}{*}{IVE-conv$\empty^{2)}$}
& \S\ref{sec:alg:1} (Algorithm 1)
& $v \rightarrow \hat{\bm{w}}_1 \rightarrow (\hat{W}_{\z}, \Omega) \rightarrow \cdots \rightarrow \hat{\bm{w}}_{K} \rightarrow (\hat{W}_{\z}, \Omega)$
& \multirow{2}{*}{$\mathrm{O}( (K + 1) L^2 FT + (K + 1) L^3 F )$}
\\
&& \S\ref{sec:alg:2} (Algorithm 2)
& $v \rightarrow G \rightarrow \bm{w}_1 \rightarrow (W_{\z}, \Omega) \rightarrow \cdots \rightarrow \bm{w}_{K} \rightarrow (W_{\z}, \Omega)$
\\
\hline
\multicolumn{5}{l}{$\empty^{1)}$
    We use the notations $\hat{W}_{\z} = [\hat{ \bm{w}}_{K + 1},\ldots,\hat{\bm{w}}_M ] \in \mathbb{C}^{(M + L) \times (M - K)}$
    and $W_{\z} = [ \bm{w}_{K + 1}, \ldots, \bm{w}_M ] \in \mathbb{C}^{M \times (M - K)}$.
}
\\
\multicolumn{5}{l}{$\empty^{2)}$
    The IVA and IVE source models can be freely changed to the ICA and ILRMA source models, and so we discuss only the IVA or IVE models.
}
\end{tabular}
}
\end{center}
\vspace{-2 mm}
\end{table*}

\subsection{Algorithm 1: Update each convolutional filter one by one}
\label{sec:alg:1}

To solve problem \eqref{problem:maxdet-conv},
we propose a cyclic BCD algorithm that updates
$\hat{\bm{w}}_1 \rightarrow (\hat{W}_{\z}, \Omega) \rightarrow \cdots \rightarrow \hat{\bm{w}}_K \rightarrow (\hat{W}_{\z}, \Omega)$ one by one by solving the following subproblems:
\begin{align}
\label{problem:ip1:wi}
\hat{\bm{w}}_i &\in \argmin_{\hat{\bm{w}}_i} \hat{g}(\hat{\bm{w}}_1,\ldots,\hat{\bm{w}}_K,\hat{W}_{\z}, \Omega, v),
\\
\label{problem:ip1:Wz}
(\hat{W}_{\z}, \Omega) &\in \argmin_{(\hat{W}_{\z},\, \Omega)} \hat{g}(\hat{\bm{w}}_1,\ldots,\hat{\bm{w}}_K,\hat{W}_{\z}, \Omega, v).
\end{align}
From the observation given in Section~\ref{sec:reduction},
these subproblems can be equivalently transformed to
\begin{align}
\label{problem:wi:alg1}
\bm{w}_i &\in \argmin_{\bm{w}_i} \bm{w}_i^h V_i \bm{w}_i - \log \det \big( W^h W \big),
\\
\label{problem:Wz:alg1}
(W_{\z}, \Omega) &\in \argmin_{(W_{\z},\, \Omega)} g_{\z}(W_{\z},\Omega),
\\
\nonumber
g_{\z}(W_{\z},\Omega) &= \trace \big( W_{\z}^h V_{\z} W_{\z} \Omega^{-1} \big) - \log \det \big( W^h W \Omega^{-1} \big) 
\end{align}
through \eqref{eq:opt:wi-wpe} and \eqref{eq:opt:Wz-wpe}, respectively.
Here, $V_i$ and $V_{\z}$ are defined by \eqref{eq:Vi}.
As shown in~\cite{ono2011auxiva}, problem \eqref{problem:wi:alg1} can be solved as
\begin{align}
\label{eq:ip1:wi:1}
\bm{u}_i &\leftarrow ( W^h V_i )^{-1} \bm{e}_i \in \mathbb{C}^M,
\\
\label{eq:ip1:wi:2}
\bm{w}_i &\leftarrow  \bm{u}_i ( \bm{u}_i^h V_i \bm{u}_i )^{-\frac{1}{2}} \in \mathbb{C}^M,
\end{align}
where $\bm{e}_i$ is the $i$-th column of $I_M$.
On the other hand, as shown in {\cite[Proposition 4]{ike2020ive}}, problem \eqref{problem:Wz:alg1} can be solved as
\begin{align}
\label{eq:ip1:Wz}
W_{\z} &\leftarrow \begin{bmatrix}
( W_{\s}^h V_{\z} E_{\s} )^{-1} (W_{\s}^h V_{\z} E_{\z} )
\\
-I_{N_{\z}}
\end{bmatrix} \in \mathbb{C}^{M \times N_{\z}},
\\
\label{eq:ip1:Omega}
\Omega &\leftarrow W_{\z}^h V_{\z} W_{\z} \in \mathcal{S}_{++}^{N_{\z}},
\end{align}
where
$W_{\s} \coloneqq [\bm{w}_1,\ldots,\bm{w}_K] \in \mathbb{C}^{M \times K}$,
$E_{\s} \in \mathbb{C}^{M \times K}$ is the first $K$ columns of $I_M$,
and $E_{\z} \in \mathbb{C}^{M \times N_{\z}}$ is the last $N_{\z}$ columns of $I_M$,
i.e., $[E_{\s}, E_{\z}] = I_M$.

\begin{remark}
The update formula for $\bm{w}_i$,
i.e., \eqref{eq:opt:wi-wpe}, \eqref{eq:Vi}, \eqref{eq:ip1:wi:1}, and \eqref{eq:ip1:wi:2},
has already been developed in our previous paper \cite{nakatani2020computationally,ike2020asj} in a different manner.
In this subsection, we reveal that it can also be developed by exploiting the stationary condition.
The efficient update formula for $\hat{W}_{\z}$,
i.e., \eqref{eq:opt:Wz-wpe}, \eqref{eq:Vi}, and \eqref{eq:ip1:Wz}, is newly developed based on the stationary Gaussian assumption of the background noises.
There is no need to update $\Omega$ as it does not affect the behavior of the algorithm.
\end{remark}

\subsection{Algorithm 2: Alternate update of WPE and ICA}
\label{sec:alg:2}

In Section~\ref{sec:model}, we recalled by~\eqref{eq:GW} that convolutional filter $\hat{W}$ can be decomposed into WPE prediction matrix $G$ and ICA separation matrix $W$.
Here, we develop a new cyclic BCD that updates 
$G \rightarrow \bm{w}_1 \rightarrow W_{\z} \rightarrow \cdots \rightarrow \bm{w}_K \rightarrow W_{\z}$ 
one by one by solving the following subproblems:
\begin{alignat}{3}
\label{problem:G}
G &\in \argmin_{G} \hat{g} (G, \bm{w}_1,\ldots,\bm{w}_K, W_{\z}, \Omega, v),
\\
\label{problem:wi}
\bm{w}_i &\in \argmin_{\bm{w}_i} \hat{g} (G, \bm{w}_1,\ldots,\bm{w}_K, W_{\z}, \Omega, v),
\\
\label{problem:Wz}
(W_{\z}, \Omega) &\in \argmin_{(W_{\z},\, \Omega)} \hat{g} (G, \bm{w}_1,\ldots,\bm{w}_K, W_{\z}, \Omega, v).
\end{alignat}

When $K = M$ and there are no noise components,
problems \eqref{problem:G} and \eqref{problem:wi} have already been discussed in \cite{kagami2018wpe-ilrma,nakatani2020computationally,yoshioka2011wpe-ica,yoshioka2012wpe-mimo}.
However, the conventional algorithms to solve \eqref{problem:G}
suffer from a huge computational cost as shown in Table~\ref{table:alg}.
We thus propose a more computationally efficient algorithm.

\subsubsection{Algorithm to solve problems \eqref{problem:wi} and \eqref{problem:Wz}}

We first explain how to solve problems \eqref{problem:wi} and \eqref{problem:Wz}.
By substituting Eq.~\eqref{eq:GW} into objective function $\hat{g}$,
these problems can be simply expressed as
problems \eqref{problem:wi:alg1} and \eqref{problem:Wz:alg1}, respectively,
except that $V_i$ is replaced by the following $V_i'$ for each $i \in \{ 1,\ldots,K, \z\}$:
\begin{align}
V_i' &= 
\begin{bmatrix} I_M \\ -G \end{bmatrix}^h
\hat{R}_i
\begin{bmatrix} I_M \\ -G \end{bmatrix}
\in \mathcal{S}_{++}^M.
\end{align}
%
Thus, in the same way as in the previous subsection,
problem \eqref{problem:wi} can be solved as \eqref{eq:ip1:wi:1}--\eqref{eq:ip1:wi:2}, where
$V_i$ is replaced by $V_i'$.
Also, problem \eqref{problem:Wz} can be solved as \eqref{eq:ip1:Wz}--\eqref{eq:ip1:Omega}, where $V_{\z}$ is replaced by $V_{\z}'$.

\subsubsection{Algorithm to solve problem \eqref{problem:G}}

We next propose an algorithm to solve \eqref{problem:G}
with less computational time complexity than conventional ones.
Every optimal $G \in \mathbb{C}^{L \times M}$ of problem \eqref{problem:G} (when $W$, $\Omega$, and $v$ are kept fixed) satisfies the stationary condition, which can be computed as
\begin{alignat}{3}
&&&&&~~ O_{L,M} = \frac{\partial \hat{g}}{\partial {G}^\ast}
= - \left. \frac{\partial \hat{g}}{\partial \bar{W}^\ast} \right|_{\bar{W} = -GW} W^h,
\\
\label{eq:G:1}
&&&\Longleftrightarrow&~&
\begin{cases}
G \bm{w}_i = \bar{R}_i^{-1} \bar{P}_i \bm{w}_i,
\quad i = 1,\ldots,K,
\\
G W_{\z} = \bar{R}_{\z}^{-1} \bar{P}_{\z} W_{\z},
\end{cases}
\\
&&&\Longleftrightarrow&~~~&
\label{eq:G}
G = 
\begin{bmatrix}
\bar{R}_1^{-1} \bar{P}_1 \bm{w}_1 \mid \cdots \mid
\bar{R}_K^{-1} \bar{P}_K \bm{w}_K \mid 
\bar{R}_{\z}^{-1} \bar{P}_{\z} W_{\z}
\end{bmatrix} W^{-1}.
\end{alignat}
Here, we used \eqref{eq:opt:wi-wpe} and \eqref{eq:opt:Wz-wpe} to derive \eqref{eq:G:1}.
Because problem \eqref{problem:G} is (strictly) convex, the update formula \eqref{eq:G} gives the (unique) global optimal solution.
The computational time complexity to calculate \eqref{eq:G} is shown in Table~\ref{table:alg}, which is much smaller than that of the conventional methods.

\section{Experiment}
\label{sec:exp}

In this experiment, we evaluated the signal extraction and runtime performance of the four methods described in Table~\ref{table:exp-alg}.
\begin{table}[t]
{
\footnotesize
\centering
\caption{Methods tested in experiment}
\label{table:exp-alg}
\vspace{-2 mm}
\begin{tabular}{c|l}
\hline
Method & \multicolumn{1}{c}{Description}
\\ \hline
IVE~\cite{scheibler2019overiva,ike2020ive} &
Identical to IVE-conv-(Alg1) with $L = 0$
\\
\hline
\multirow{2}{*}{IVA-conv~\cite{nakatani2020computationally}} &
An integration of WPE and IVA, which is identical to
\\
& IVE-conv-(Alg1) with $K = M$, $D_1 = 2$, and $D_2 = 5$.
\\
\hline
IVE-conv-(Alg1) & IVE-conv with $D_1 = 2$ and $D_2 = 5$ using Algorithm 1.
\\
\hline
\multirow{2}{*}{IVE-conv-(Alg2)} & IVE-conv with $D_1 = 2$ and $D_2 = 5$ using Algorithm 2.
\\
& For every five updates to $v$ and $W$, we updated $G$ once.
\\
\hline
\end{tabular}
}
\vspace{-0 mm}
\end{table}

\textit{Dataset}:
We generated synthesized convolutive noisy mixtures of two speech signals.
We obtained speech signals from the test set of the TIMIT corpus~\cite{timit}
and concatenated them so that the length of each signal exceeded 10 seconds.
We obtained point-source noise signals recorded in a cafe (\textsf{CAF}) and a pedestrian area (\textsf{PED})
from the third `CHiME' Speech Separation and Recognition Challenge (CHiME-3)~\cite{chime3}.
Note that the noise signals are nonstationary, but are considered to be more stationary than speech signals.
We obtained RIR data recorded in room \textsf{OFC} from the RWCP Sound Scene Database in Real Acoustical Environments~\cite{rwcp}.
The reverberation time ($\mathrm{RT}_{60}$) of room \textsf{OFC} is 780 ms.

The generated mixtures consisted of $K = 2$ speech signals and six noise signals randomly chosen from the above dataset.
The SNR of each mixture was adjusted to
$\mathrm{SNR} = 10 \log_{10} \frac{
    (\lambda_1^{(\mathrm{s})} + \lambda_2^{(\mathrm{s})}) / 2
}{
    \lambda_1^{(\mathrm{n})} + \cdots + \lambda_6^{(\mathrm{n})}
} = 5$ or 10 [dB],
where
$\lambda_i^{(\mathrm{s})}$ 
and
$\lambda_j^{(\mathrm{n})}$ 
denote the sample variances of the $i$-th speech signal ($i = 1,2$)
and the $j$-th noise singal ($j = 1,\ldots,6$).

\textit{Criteria}:
Using \textit{museval}~\cite{museval},
we measured the signal-to-distortion ratio (SDR)~\cite{vincent2006sdr}
between the separated and oracle spatial images of the speech signals at the first microphone.
The oracle spatial images were obtained by truncating the RIRs at 32 ms (i.e., the points after 32 ms were replaced by 0)
and convolving them with the speech signals.

\textit{Conditions}: 
The sampling rate was 16 kHz,
the frame length was 2048 (128 ms),
and the frame shift was 512 (32 ms).

\textit{Initialization}:
For all methods, we initialized the convolutional filter as $W(f) = -I_M$ and $\bar{W}(f) = G(f) = O$,
and then updated $W_{\z}(f)$ once using~\eqref{eq:ip1:Wz} before the optimization.

\begin{figure}[t]
\centering
\begin{subfigure}[b]{\linewidth}
\includegraphics[width=\linewidth]{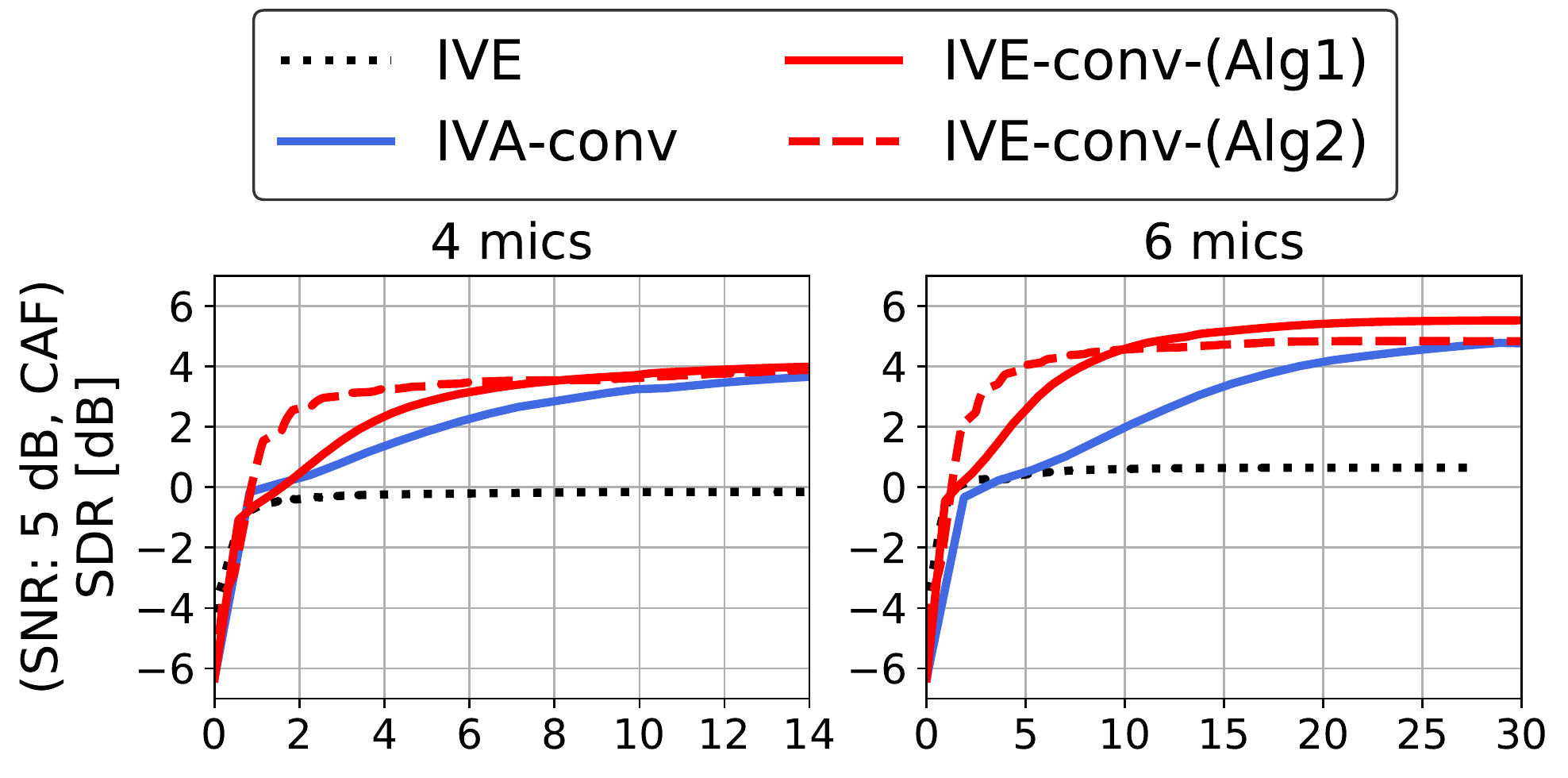}
\end{subfigure}

\begin{subfigure}[b]{\linewidth}
\includegraphics[width=\linewidth]{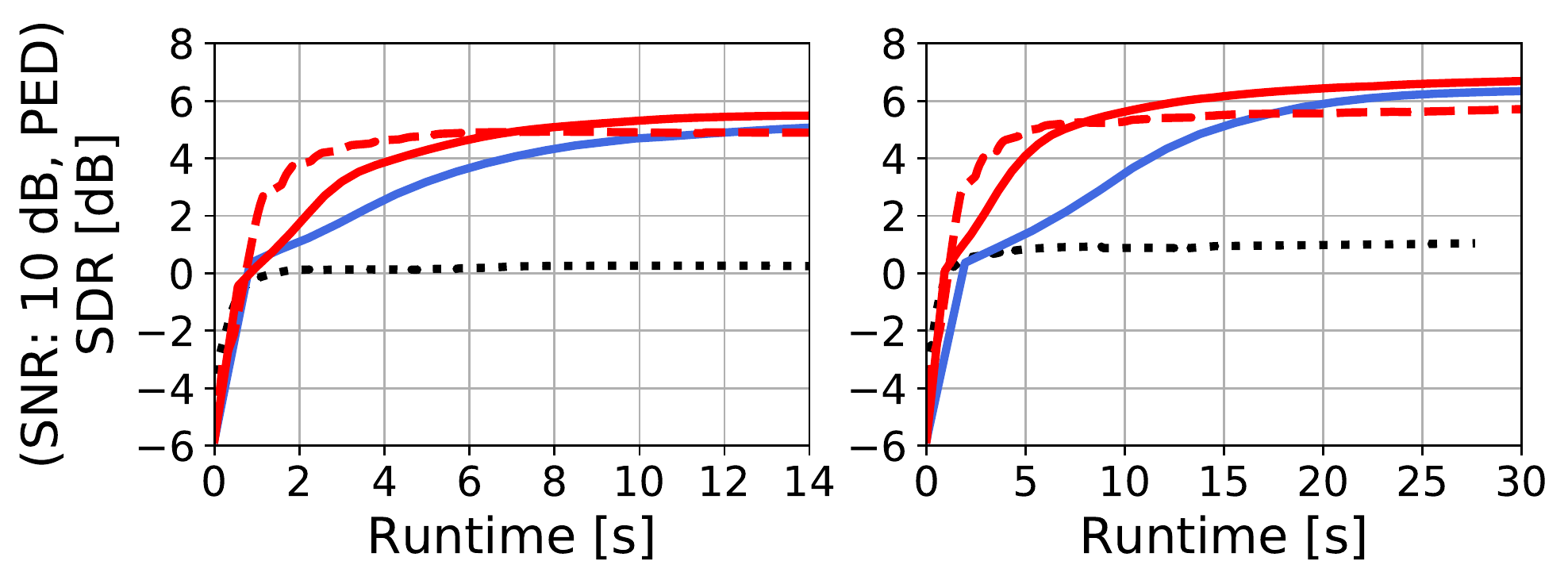}
\end{subfigure}
\vspace{-6 mm}
\caption{
SDR [dB] performance as a function of runtime.
The noise condition was \textsf{CAF} with $\mathrm{SNR} = 5$ [dB] (top) or \textsf{PED} with $\mathrm{SNR} = 10$ [dB] (bottom),
and the number of microphones was $M = 4$ (left) or 6 (right).
Results shown were averaged over 50 mixtures
and obtained by running the algorithms on a PC with ``Intel(R) Core(TM) i7-7820 CPU @ 3.60 GHz'' using a single thread.
The average length of the mixture signals is 12.51 sec.
The separated spatial image was obtained by $(W(f)^{-h} \bm{e}_i) (\hat{\bm{w}}_i(f)^h \hat{\bm{x}}(f,t)) \in \mathbb{C}^M$
for each source $i = 1,2$.
}
\label{fig:SDR}
\end{figure}

\subsection{Experimental results}
Figure~\ref{fig:SDR} shows the convergence of the SDR when each method was applied.
Compared to IVE, which does not handle reverberation, both IVA-conv and IVE-conv showed the higher SDRs.
Although the SDR performance at the convergence points is comparable,
the convergence of the proposed IVE-conv was much faster than that of IVA-conv since the computational cost to update $\hat{W}_{\z}$ is much lower.
This fast convergence behavior is important in practice,
since using more microphones can improve the SDR at the expense of increased runtime as observed in Fig.~\ref{fig:SDR}.
IVE-conv-(Alg2) converged faster than IVE-conv-(Alg1), but gave a slightly lower SDR.

\section{Conclusion}
To achieve joint source separation and dereverberation with a small computational cost,
we proposed IVE-conv, which is an integration of IVE and WPE.
We also developed two efficient BCD algorithms for optimizing IVE-conv.
The experimental results showed that IVE-conv yields significantly faster convergence than the integration of IVA and WPE while maintaining its separation performance.

\vfill\pagebreak
\bibliographystyle{IEEEtran}
\bibliography{refs}
\end{document}